# Enhanced distribution of molecules in the brain due to oscillations of the interstitial flow

Raghu Raghavan[a*], and Richard D. Penn[b]

[a*] *Corresponding Author:* Raghu Raghavan, Sonovance, Inc., Baltimore, MD

*Email address*: raghu@sonovance.com

[b] Richard D. Penn MD, UIC Biomedical Engineering, Chicago Il

**ABSTRACT**

Background: The movement of molecules in the interstitial spaces of the brain is most often viewed as being governed primarily by passive Brownian diffusion with a slow advective component. MRI measurements from a decade-old study of the physical properties of brain tissue observed a dynamic, pulsating fluid flow attributed to the cardiac cycle. The effects of this cyclic flow pattern on the spatial distribution of molecules in the brain are modeled in this paper.

Methods: The effects of oscillatory flow on the dispersion or volumetric transmission of a molecule that is advected by this flow is modeled by a mechanism hitherto neglected in the literature. The MRI measurements of water flow in and out of brain tissue due to cardiac pulsations have been used in a simplified model to quantify the distance a molecule would move when carried by this flow. An oscillatory random walk model is used to estimate the spread or effective diffusivity due to the oscillatory advection. Then, respiration effects are also estimated and the additional dispersion of molecules due to this are calculated in our model.

Results: Our model indicates that the observed oscillatory flow in the interstitial spaces due to cardiac as well as respiratory pulsatility can induce an effective diffusivity when the spread of the molecule is observed over times long compared with a cycle of the oscillation. This would help explain the high-volume transmission within the interstitium or brain parenchyma found in MRI measurements of a marker infused into the cerebrospinal fluid in human subjects that is well above what would be expected.

Conclusions: Interstitial spaces should be viewed as a region of dynamic oscillatory flow driven by cardiac and respiratory cycles. This oscillatory flow could result in a significant dispersion of molecules and explain the higher-than-expected effective diffusion suggested in human studies. It may be possible to augment or slow this flow and concomitant spread by applying external forces.

*Abbreviations*: CSF (cerebrospinal fluid), MI (mechanical index), MRI (magnetic resonance imaging), SAS (in this paper, without a qualifier it denotes the cranial subarachnoid space).

# 1 Introduction

## *1.1 Plan of the paper*

The principal result in this paper is an endogenous mechanism, overlooked hitherto as far as we are aware, that increases the spread of a molecule within the interstitial spaces of the brain. The extent or effectiveness of the spread is co-determined by the cardiac and respiratory pulsations which result in oscillatory fluid flow in the interstitium and the spatial randomness (and spacing) of the branchings in the interstitial pathways. The motivation for this paper was a remark in a study of the spread of an MRI contrast reagent in (Vinje, Zapf et al. 2023) which serves as a surrogate tracer for some therapeutic molecules of interest in the treatment of brain disease cannot be explained by diffusion alone.

The plan of the paper is as follows: The body of the paper (Section 2, **Methods**) begins by developing a result in random walk. A standard elementary result in random walk in three dimensions is that if a particle takes a step of length $L$ in time $\tau$, randomly oriented in space with respect to the previous step, then at the end of $N$ steps, its mean distance (over samples of such walks) from the starting point is 0, and its mean squared distance is $L^2 N$. As is well known, this is classical diffusion when taken to the continuum limit. It is shown first from numerical study that if each step is random, *but constrained to have a negative projection onto the previous step's direction*, then diffusion still occurs though with a reduced diffusivity, $1/3^{\text{rd}}$ of the conventional random walk's. For the purposes of this paper this is referred to as an "oscillatory random walk". Section 2.2 then argues that the interstitial pathways are sufficiently randomly branched to constitute pathways for a random walk. Then, provided a certain constraint is met, oscillatory fluid flows in the brain that can advect particles will result in the oscillatory random walk referred to. This is the proposed mechanism that enhances particle spread. The final piece of the argument of course is the existence of such fluid flows. The evidence for this from an experiment followed by a model calculation is presented in Section 2.3. The **Results** Section 3 ties together the calculations from the previous section to make the case for the mechanism and its magnitude in the human brain.  An important aspect of this mechanism is that it is independent both of the magnitude of the particle diffusivity and of particle size, within limits that will be made clear in the section.

There is a vast literature attempting to understand perivascular and interstitial fluid flows in the brain along with implications for drug delivery.  In order not to impede the development, we a discussion of these is in the final Section 4, **Discussion**, which begins with the motivating remark of (Vinje, Zapf et al. 2023) mentioned earlier and then proceeds to discuss other literature which is complementary to our work.   In conclusion, some speculations on the possibility of enhancing this mechanism to distribute drugs in the brain with artificial means is offered.

## 2 Methods

### *2.1 Random walk and interstitial pathways*

The central result of this paper and its implications, developed in subsequent sections to apply to the brain interstitium, is illustrated with some simplified models.  In Figure 1, a two-dimensional square lattice (which has the virtue of allowing an exact calculation, in fact in any dimension) is shown, with the interstitial pathways in white being arranged in a “Manhattan”-type grid with square blocks representing the cellular obstructions.  Figure 1a shows a particle

**Figure 1**

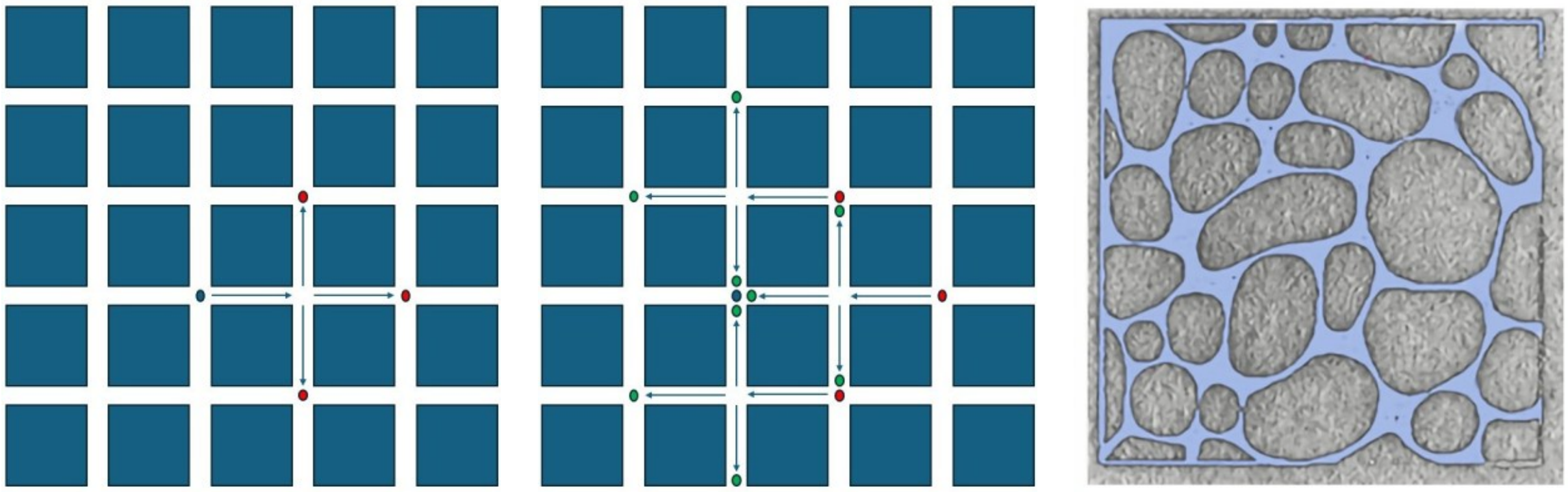

(a) (b) (c)

advected, beginning with the black dot.  In the first half-cycle (Figure 1a) the particle is advected to a (taxicab) distance of 2.   The key assumption made is that at any branch, the fluid flow and therefore the advected particle can choose any of the branches as long as it is not oppositely directed to the pressure driving the flow. The particular assumptions for the walk on this lattice is clearly shown in Figure 1a: the particle will arrive at any of the positions marked in red with equal probability. Following the same assumption, *mutatis mutandis*, for the second half-cycle results in the particle arriving at the green dots: each dot representing a contribution of equal probability. It is obvious that the particle has not returned to the origin: a cloud of particles will spread. The mean distance is of course zero, the center of mass of the dots is the origin. The picture can be converted into an exact, if elementary, result: The mean square distance traversed in one cycle, with obstacles of unit length, and distances traversed twice that, is 16/9 in the two-dimensional model and 48/25 in the three-dimensional, both are close to 2. If one cycle is about one second (as for the cardiac cycle) and the distance traversed is 20 microns this result in a mean square distance in one second of 100 $\mu m^2$. The effective diffusivity of the particle is therefore augmented by 100/6 ~ 16 $\mu m^2$/second using the result that the mean square distance is 6 D t for a Wiener or diffusive process, where t is the time and D the diffusivity. Two other characteristics of this dispersion may be mentioned: (i) Namely, it is entirely independent of molecular diffusivity or size of the molecule, subject of course to the constraint that it can be advected freely by the fluid. (However, molecular diffusivity will aid the process of the particle being able to choose a different pathway in the second half-cycle from whence it came in the first.) (ii) The molecular diffusivity or Brownian motion may be sufficient to carry the particle past a branch but by definition, this will not aid the process mentioned. Advection by fluid is necessary. Finally, the model and result just discussed have been used only for illustrative purposes and the resulting augmentation of diffusivity is inadequate to explain the experimental results to be described later.

*It is a central assumption of the model proposed that the trajectory the particles choose is random as long as it has a component in the direction of the advective pressure gradient.* Ultimately, the validity of the assumption will be decided by more careful measurement and more detailed computer simulations than are shown in this paper. However, the assumption is not unreasonable. The interstitium is host to many macromolecules (obstacles in the path of the particles). These obstacles moreover are moving around due to the motion of the tissue and of the fluid, in addition to any Brownian motion or diffusion. The particles (which may also undergo Brownian motion) returning to the branch have no reason to retrace the path they took in

the first half of the cycle. Thus, *the motion of the individual molecules can be irreversible even though a macroscopic treatment of the Stokes flow carrying it is not*. Moreover, the fluid is not oscillating in any fixed direction (see for example the vector fields plotted in (Heukensfeldt Jansen, Abad et al. 2025)), so that the return path of the molecule in the second half-cycle may easily take a different branch in the network of interstitial pathways. Figure 1(c) is an illustration of a planar section of more complex three-dimensional channels for interstitial or extracellular fluid flow. The fact that random pathways effect a diffusive spread of the molecule has been well known to geophysicists for a long time. Jacob Bear (Bear 2013) refers to this as mechanical dispersion which is a term that seems to have fallen out of favor. While treatments of this due to steady advection have also been known for a long time, we are not aware of treatments due to oscillatory flow. We therefore carry out one more calculation by going to the continuum limit in three dimensions.

We will assume that the pathways at a branch are essentially randomly directed in space. If at each step (= half-cycle) the path chooses a branch randomly, we have a standard random walk in space and its well-known limit as a diffusion process in the limit of large time. To account for the oscillatory walk, we have used the following simplified model. We generate a random direction $w_n$, $n = 1, 2, \ldots$, where $w_n$ is the $n$'th sample of a unit vector chosen to be distributed uniformly on the surface of a sphere with respect to the area measure but subject to the constraint $w_n \bullet w_{n+1} < 0$. So, each direction in three-dimensional space is first randomly chosen. Then, if it has a positive projection along the direction of the previous unit vector, we flip its sign. This is, admittedly, simplified but arguably captures an essential feature of oscillatory transport. Then a sum of $T$ steps of the resulting position of a point is computed and averaged over a large number of runs. The numerical evidence indicates that the mean square distance (and, hence, effective diffusivity) is 1/3 that of the unconstrained random walk, as we will discuss. As is well known, the mean square distance is equal to the number of steps $N$ for unit length steps or $Nd^2$ for steps of length $d$. In any case, we have computed the mean square distance as a function of the number of steps, as we now describe. The calculations in our model *do assume* that the fluid velocity is the velocity of advection of the molecule. Certainly, for small-molecule gadolinium – based contrast reagents, this is the case (Brady, Raghavan et al. 2020). If the advection of a particle in interstitial space is retarded due to its size, then that will affect all calculations and we would have to substitute the particle velocity instead of the fluid velocity, as we have done here. In the literature on brain transport, this is frequently described via a fudge factor, known as the retardation factor, that multiplies the fluid velocity. However, we shall not be concerned with such aspects here.

**Figure 2**

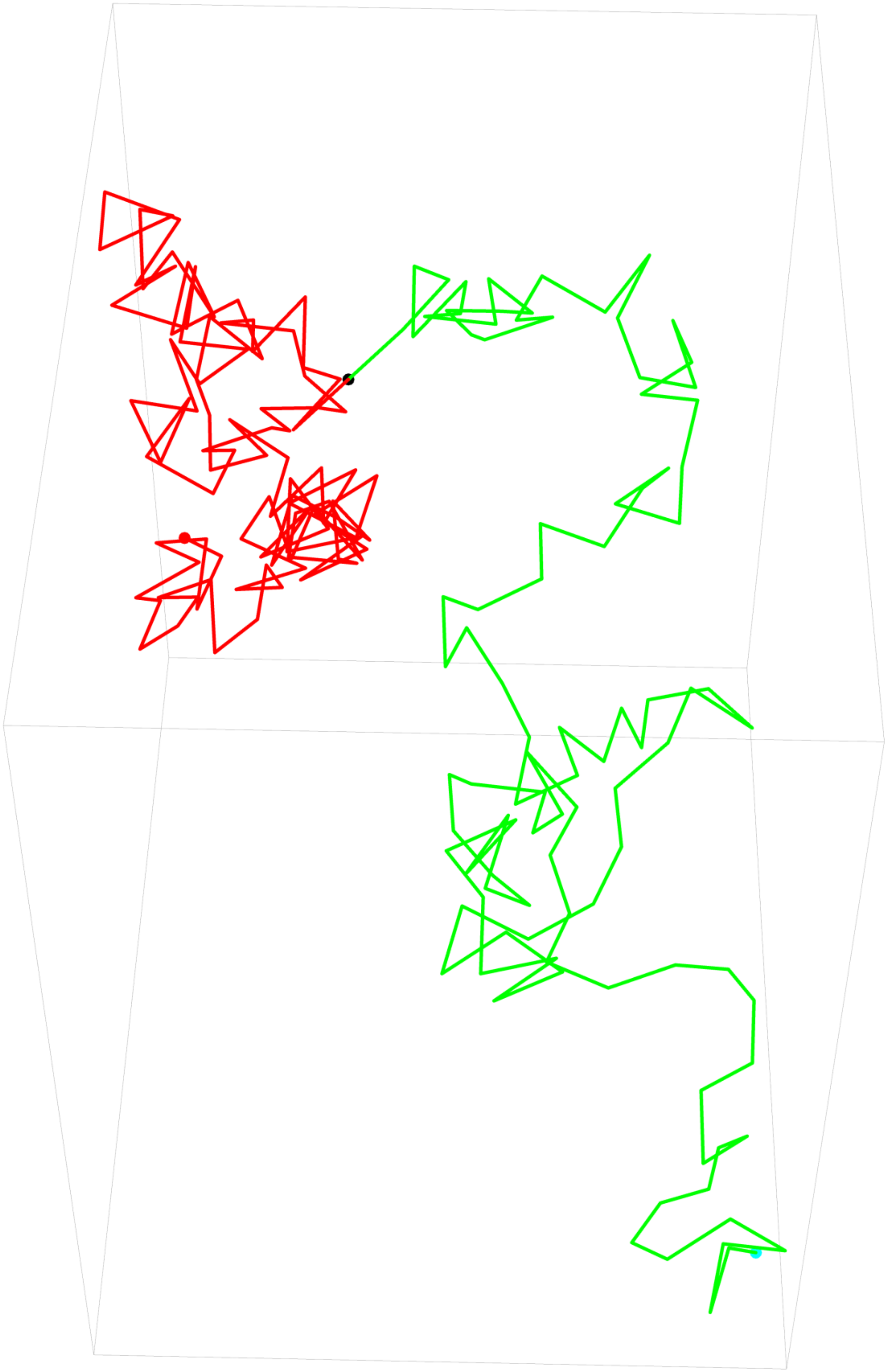

Figure 2 shows 100 steps of both a normal and an oscillatory walk in three dimensions. Both walks begin at the same point (the black sphere). The normal random walk is shown in green and ends at the blue sphere, while the oscillatory walk in red begins with the same initial step (which overwrites the green step) and ends at the red sphere. It is clear that the oscillatory walk (in red) has not wandered as far as the normal walk. It is a standard and elementary result that a normal random walk, which is the sum of $N$ unit length vectors, converges in probability to a vector with a Gaussian distribution with mean zero and variance $N$. The numerical evidence seems to indicate that the oscillatory walks also converge to a normal distribution despite the short-term memory and that the variance is 1/3 that of the usual random walk. We have not attempted to find a proof of these statements, but we have simulated up to 10,000 steps of a walk with 4000 runs. For shorter walks, we took the number of runs up to 20,000 and confirmed convergence of the mean square distance (the variance of the distribution) obtained with fewer samples. Note that 1000 steps would be about 10 minutes of a cardiac cycle (assumed 1 per second) and about 2 minutes of a respiratory cycle. Thus, the longest simulation we have done would be just under 2 hours of a cardiac cycle and 20 minutes of a respiratory cycle. This is a sufficient time within which to observe dispersive spread of a cloud of molecules deposited somewhere in the interstitium.

This difference in the distribution for the oscillatory walk is preserved on taking the average and as stated, it has (as far as the numerical evidence can take us) 1/3 of the variance of a walk where the directions are random with no constraint or correlation. The results of our calculations are shown in Figure 3: the size of the dots is larger than the variance of the results. For the standard walk, the mean square distance is equal to the number of steps, as it should. For the oscillatory walk, the numbers shown in the figure yield a mean square distance about 1/3 of the number of steps. As mentioned, we have not attempted to prove this result analytically so we do not know if the number should be exactly 1/3. The factor will, obviously, be lower in two dimensions and is zero in one dimension.

**Figure 3**

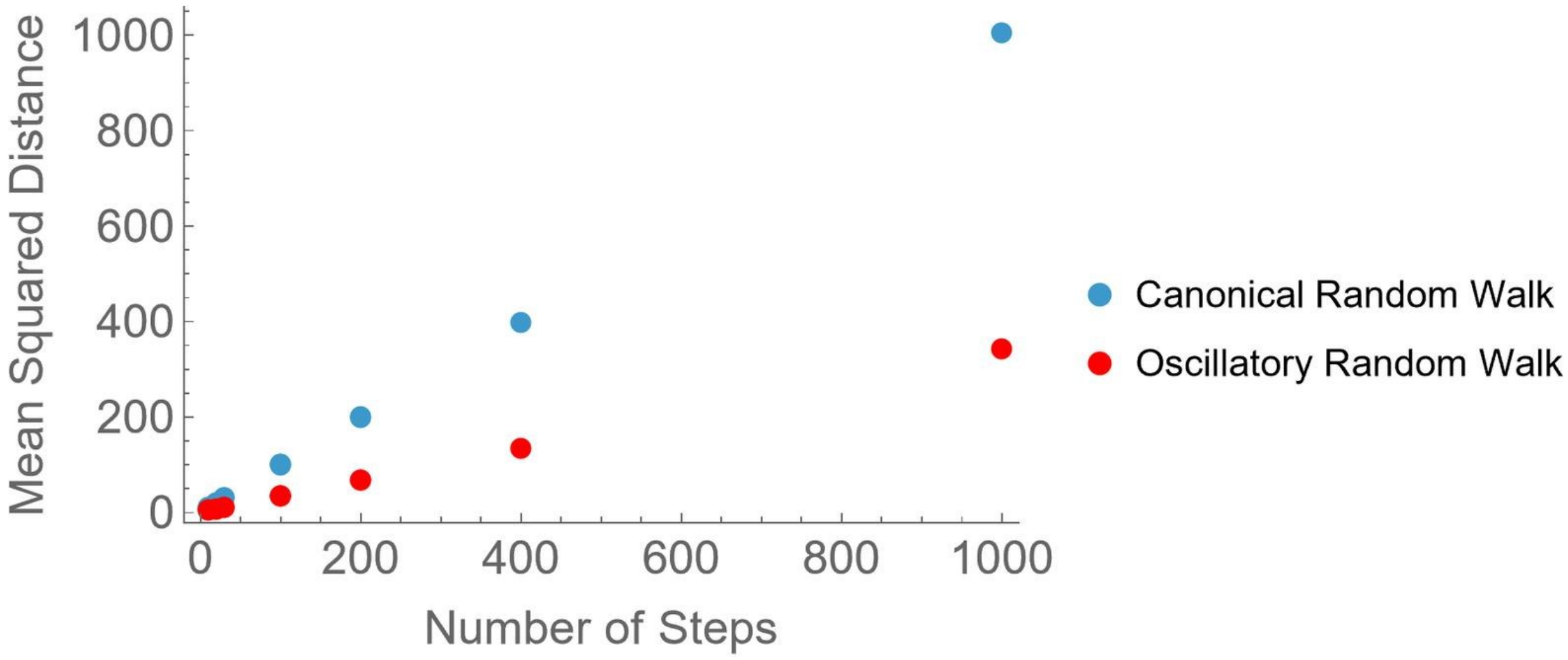

It may be noted that the cubic lattice gives a mean squared distance of ½ (more exactly, 12/25 as seen above) in one cycle with a step of unit length, while the continuum yield 1/3 under the same assumptions. For $N$ steps each of length $L$, we multiply the fraction by $L^2 N$.

*2.2 Brain interstitial pathways and oscillatory flows*

If the assumptions above are made then, to make contact with data for the brain, the advective speeds and distances in a half-cycle for brain interstitial flows need to be obtained. We rely on an experimental result (Section 2.2.1) which is then used in a model (Section 2.2.2) to obtain estimates for the flow speeds and distances. *The purpose of this subsection is to provide a model for the distances to be used in the random walk model, and not to provide the actual trajectories of fluid flow as discussed above.*

2.2.1 Input data on oscillatory interstitial flow

Our experimental input is from the paper by Hirsch et al. that determined the outward and inward flow of water in multiple regions of the brain related to the cardiac pulse (Hirsch, Klatt et al. 2013). By timing the measurements to the cardiac cycle, the oscillatory flow into and out of the regions of interest due to cardiac pulsatility can be quantified. The divergent flow due to compression represents "the interstitial fluid being squeezed out of the region … by the arterial pulsation" (Hirsch, Klatt et al. 2013). This substantial oscillatory flow along the cells through the complex, branching fluid spaces outside the cells means that the effective diffusion of molecules in these spaces must be seen in the context of these oscillations. These data on the divergence div(***v***) of the interstitial flow velocities from MRI measurements is presented in Hirsch *et al.*, (Hirsch, Klatt et al. 2013) shown as Figure 4 and redrawn from their Figure 5(c) (*loc cit.*). Their equation (5) and the incompressibility property of the tissue, as in their equation 6(c), mean that the figure represents the divergence of the Darcy flux: the integral of the quantity plotted at a given time over a specified volume of tissue, which, in their case, is the flux of fluid through the surface of the volume of the region of interest drawn on the MRI slice. (Darcy flux has the dimensions of velocity but is the volume of fluid crossing a unit area in tissue: the orientation of the area defines the direction of the flux.) For the purposes of our calculation, we have scaled the given original curve by a factor of 5 based on the finding that the volume of interstitial fluid is 20% of the total volume of tissue (Syková and Nicholson 2008). We denote this interstitial volume fraction as $\phi$. The reason for this is that we will need to follow the advected particle in the interstitial spaces, and the interstitial fluid velocity is a factor $1/\phi$ times the Darcy velocity; see (Morrison, Laske et al. 1994) as well as the discussion following equation (1) below. We note that in pathological conditions the space can vary considerably in size from the normal value of 0.2 to double that in brain tumors or half with ischemia. (See Table 7 in (Syková and Nicholson 2008).) More recent studies have also looked at sleep, aesthesia, and aging conditions and demonstrated significant changes in the size of the interstitial spaces. Rather than get into arguments about these experiments, we have chosen to note a variation in size and remark on how that would affect our calculations of dispersion in the Discussion section.

The figure is plotted as having a zero mean, since the flow in and the flow out of the tissue are equal. The result in Figure 4 shows the oscillatory divergence of the relative interstitial velocity of fluid into and out of the tissue as a function of phase in the cardiac cycle. In the absence of further information, this divergence is assumed uniform throughout the parenchyma. (In the Discussion section below, we comment on this assumption.)

**Figure 4**

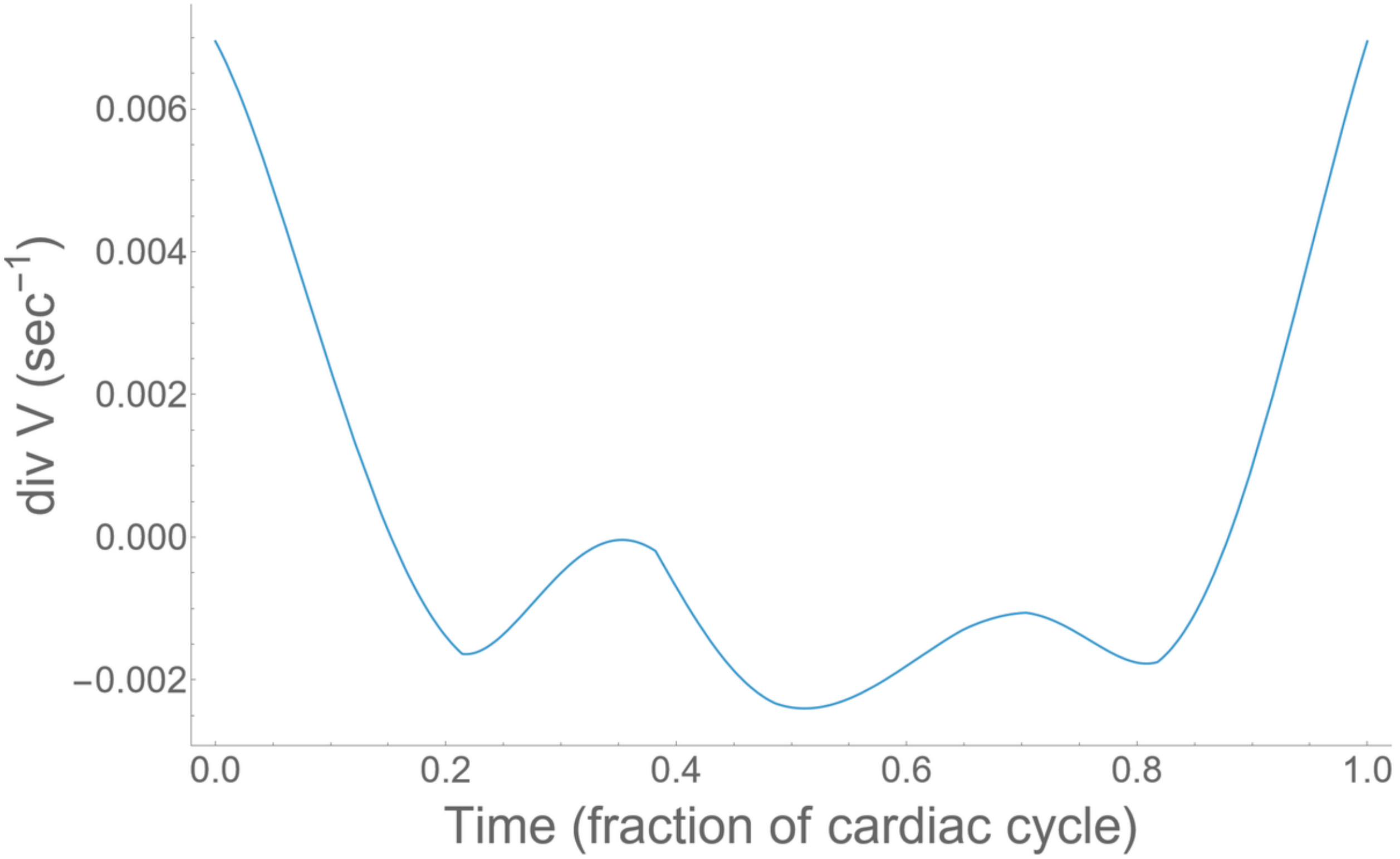

**Extrapolating to the respiratory cycle**.

The Hirsch paper provides data on the effect of abdominal pressure on div(***v***), which clearly implies respiratory effects. However, they did not time div(***v***) with the respiratory cycles. With improved technology, new strain measurements are now available (Sloots, Biessels et al. 2020, Sloots, Biessels et al. 2021). In particular, these authors re-examined the cardiac-induced parenchymal strain measurements of earlier work and also measured the respiratory-induced strain. Unfortunately, the paper does not provide a time course of the divergence of the Darcy flux over a respiratory cycle, which is what we would need for a computation. Nevertheless, their cardiac-induced strain values are, indeed, comparable with those reported by Hirsch, and in fact, even slightly larger in the basal ganglia, supporting our use of their waveform (Hirsch, Klatt et al. 2013). The Sloots group also found that respiratory-induced strain was about 1/3 of the cardiac strain: “Furthermore, heartbeat-induced volumetric strain was about three times larger than respiration-induced volumetric strain” (Sloots, Biessels et al. 2021). This number is used in the calculations below.

*2.3 Estimating the interstitial flow velocities and excursion distance of advected particles*

2.3.1 Brain model

The brain model used is shown in Figure 5, the dimensions being derived from Table I of (Lüders, Steinmetz et al. 2002). The model is meant to be that of a region bounded by a sphere, not a circle. However, at the outset and to avoid any misunderstanding, it should be stated that the geometry of the different regions (gray and white matter) has absolutely no impact on the calculations due to assuming full spherical symmetry and no variation of the interstitial volume fraction in this model.

**Figure 5**

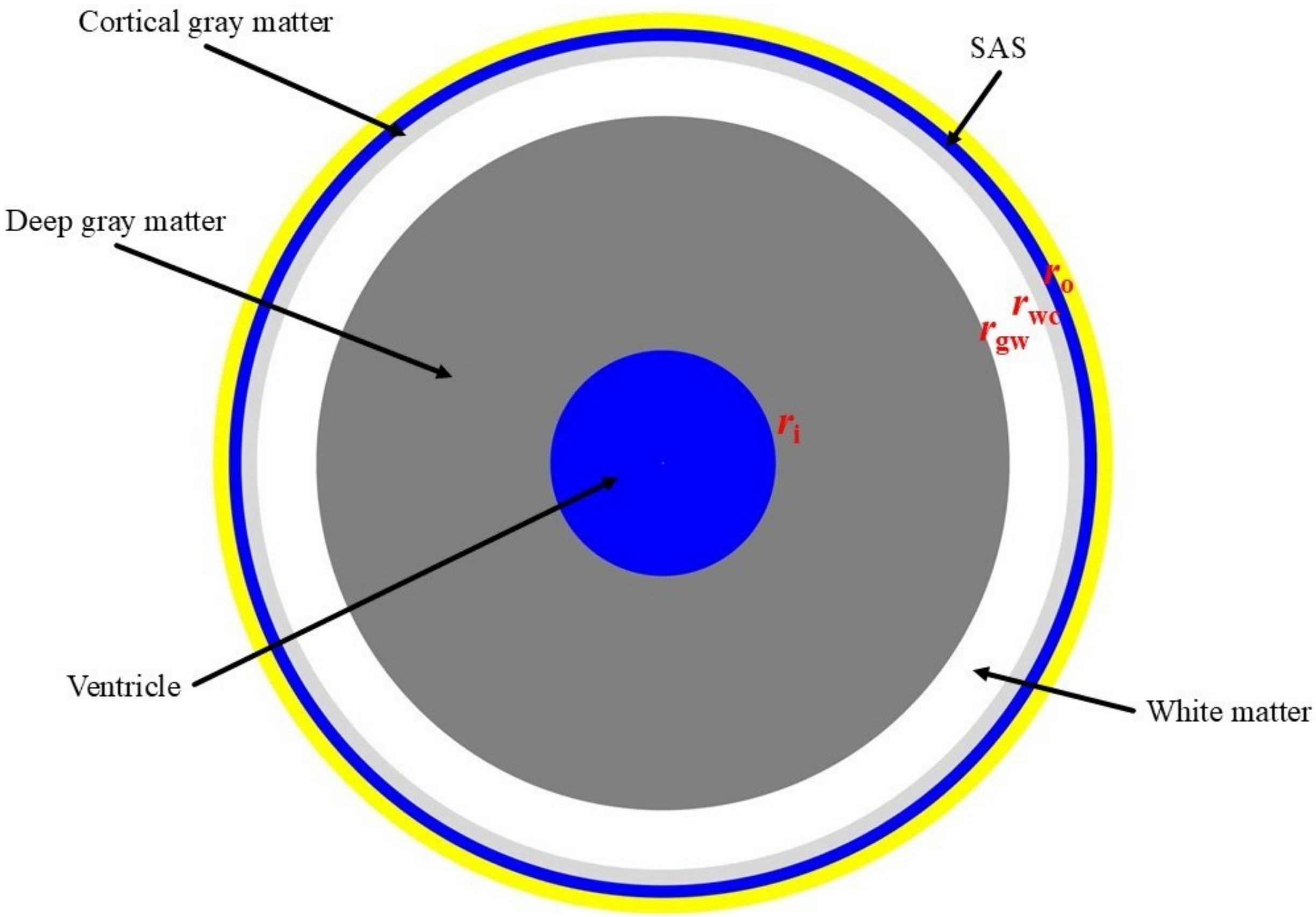

The numbers listed in the table for $r_i$, $r_{gw}$, $r_{wc}$, and $r_o$ are the radii at the ventricle/inner gray matter boundary, the inner gray matter/white matter boundary, the white matter/outer cortical boundary, and the cortical/cranial subarachnoid boundary, respectively. Table 1 gives the values of these radii and the corresponding (approximate) volumes of the spherical shells. These numbers are very approximate, but see (Im, Lee et al. 2008), which gives a variation from 1.15 to 1.98, with a mean of $1.5 \times 10^6$, mm³ for the total volume of the brain.

**Table 1**

| Region | Color in the figure of the brain model | Radius (cm) | Volume (cc) |
|---|---|---|---|
| Lateral ventricles | Blue | $0 < r < 1.8$ | 24.5 |
| Deep gray matter | Dark gray | $1.8 < r < 5.53$ | 684 |
| White matter | White | $5.53 < r < 6.475$ | 429 |
| Cortical gray matter | Gray | $6.475 < r < 6.725$ | 268 |
| CSF/skull | Blue /yellow | $6.725 < r < 7.175$ | Total parenchymal volume = 1766 |

### 2.3.2 Model equations and solution for interstitial velocities and advected distances

The key question is this: Does the measured water flux, the MRI-derived divergence, cause enough flow in the interstitial space of the brain to result in morphological dispersion? In turn, the key to understanding whether morphological dispersion could be important in the interstitial spaces is knowing the distance traversed by a particle of neutral buoyancy and free to advance by advective fluid movement. Thus, to answer the question of interest, namely the distribution of the particle or molecule, the fluid velocities obtainable from the measured fluxes are computed. These then are used to compute the transport of solute that is advected with these velocities. Once the velocity fields are known, the distance an advected particle is transported is obtained for each scenario by simply integrating the velocity of the fluid over a half-period. Subsequently, we add the potential effects of respiration on these movements. Finally, we use an elementary random walk model to infer the dispersivities or the effective diffusion coefficients.

To calculate the velocity fields under the basic assumptions, namely the data on the divergence from Hirsch et al. (2013), we use the equation for Stokes flow upscaled for a porous medium, *ie.*, Darcy's law. Due to the assumption of full spherical symmetry, the calculations are elementary.

The equations for the flow within parenchyma, including the oscillatory component (or, in this case, *only* the oscillatory component, since we will confine ourselves to linear flow), are written here for full spherical symmetry with only a radial velocity denoted $v$, and a radial derivative denoted by a prime. The radial coordinate is $r$, and full spherical symmetry is assumed. We use Darcy's law denoting the hydraulic conductivity by $K$ (here a fully isotropic, homogeneous tensor, thus a single number) along with the fluid continuity equation:

$$\mathbf{v} = -K\nabla p; \quad \mathrm{div}\,\mathbf{v} =: q[t] \tag{1}$$

In equation (1), $\mathbf{v}$ denotes the fluid velocity (assumed to have only a radial component in our spherically symmetric model) and $p$ is the fluid pressure. $q[t]$ is the *average measured* divergence in parenchyma, as given in Hirsch et al. (2013), and is a known function of $t$. We note that although the pressure and the velocity are time-dependent, the acceleration term can be safely neglected because the hydraulic conductivity of tissue is so small. Estimates vary greatly in the literature, but they are at most $10^{-10}$ m/s per Pa/m pressure gradient(Morrison et al. 1994). Care must be taken because we will use the word "velocity" to mean the interstitial velocity and not the Darcy velocity: see (Morrison et al. 1994) for a discussion of this point. The Darcy flux refers to the volume of fluid flowing past a unit area of *tissue*, while the interstitial flux refers to the flux past a unit area of the interstitial fluid space. The volume fraction of the interstitial fluid, a factor of $\phi$, relates the two. We, thus, multiply the measured divergence by $1/\phi$, where $\phi$ is the interstitial volume fraction as mentioned in Section 2.1. We will soon see that the variation in $K$ that depends on this choice is entirely irrelevant.

To solve for $(v, p)$, where $v$ is a scalar function since we assume the velocity is radial, we need the boundary condition for the pressure $p$. Let $\Delta p[t]$ be the difference in pressure at each instant of time between the inner boundary of the parenchyma at radius $r_i$ and the outer boundary at $r_o$. Then the solution for the speed (radial velocity component) is

$$v[r,t] = q[t]\left(\frac{r}{3} - \frac{A}{r^2}\right) + K\frac{B}{r^2}\Delta p[t] \tag{2}$$

where

$$A = \frac{r_i r_o}{6}\left(r_i + r_o\right); \quad B = \frac{r_i r_o}{r_o - r_i} \tag{3}$$

(The $r$'s in equation (3) refer to the inner and outer radii, see Figure 3.) The term in the velocity proportional to $q[t]$ may be seen to satisfy the divergence condition, obtained, as we have emphasized, from (Hirsch, Klatt et al. 2013), and yields a condition of zero pressure difference between the boundaries. The term proportional to the hydraulic conductivity $K$ has zero divergence and satisfies the condition that the pressure difference at the boundaries is $\Delta p[t]$. *However, we may immediately discard this term*. From (Linninger, Xenos et al. 2009) – see for example their Figure 5(c) – we see that the transmantle pressure difference is, at most, of the order of 100 Pa. Now it may be argued that this is not necessarily the pressure differences in the sub-pial parenchyma, but since the pia is very porous ( (Betsholtz, Engelhardt et al. 2024), paragraph on *Pial cell continuity supported by adherens junctions* and references therein), it is difficult to envisage much larger pressure differences. Furthermore, the constant $B$ is much less than 1 m (2.46 cm with our choice for the radii). Since, as we have said, the hydraulic conductivity is no larger than $10^{-10}$ in SI units ($m^4/N$-sec), we can completely disregard the second term.

Thus, to calculate the position $R$ of an advected particle of neutral buoyancy starting at $r = a$ at time 0, we need to solve $dr/dt = q[t](r/3 - A/r^2)$, which is readily done by separating the variables. We must have

$$\int_{r_i}^{r_o} \frac{1}{\frac{r}{3} - \frac{A}{r^2}}\, dr = \log \frac{R^3 - A}{a^3 - A} \tag{4}$$

equal to

$$\int_0^t q[\tau]\, d\tau =: g[t] \tag{5}$$

which yields

$$R[a,t] = \sqrt[3]{A + \left(a^3 - A\right) e^{g[t]}} \tag{6}$$

Thus, a particle dropped at place $r = a$ at time 0 will have moved to $R[a, t] - a$ at time $t$ within a cycle.

**Addition of respiratory-induced flow effects.**

To model the respiratory effects, we used the cardiac waveform over 1 s and expanded it by five times to approximately the respiratory cycle. We also lowered its magnitude by 1/3, as Sloots's measurements of the elasticity suggest (Sloots, Biessels et al. 2021). More precisely, let $q[t]$ be the divergence waveform due to the cardiac cycle, with period here taken to be 1 s.

Assume the respiratory waveform has a period of 5 s. Then the waveform for the respiratory case with subscript R is set in our computations to be

$$q_{\mathrm{R}}[t] = \frac{1}{3} q[t / 5] \tag{7}$$

## 3 Results

The calculations of the previous section yield the results shown in this section. First, the excursion distances obtained from the model in Section 2.3 are plotted, and these are then input into the random walk in the continuum model of Section 2.1. We use the continuum limit but, given the uncertainties in the calculation, it makes no difference if we were to use the cubic lattice limit.

### *3.1 Excursion distances due to cardiac and respiratory pulsations*

Figure 6 plots equation (6) at various places. Figure 6(a) shows equation (6) as is, meaning only the cardiac-driven distances are shown. The advected particle is placed near the inner boundary (blue curve), at the midpoint between the inner boundary and outer boundary (green), and finally near the outer boundary (red). The excursions of the order of 5–10 μm. Figure 6(b) follows the same scheme, but this time for the respiratory-driven cycle, as explained above. We emphasize again that these curves are not independent of those in Figure 6(a): as above, the curve is simply a scaled in time and magnitude version of the cardiac curve. Now the upper end of the excursions reaches up to 20 μm.

**Figure 6**

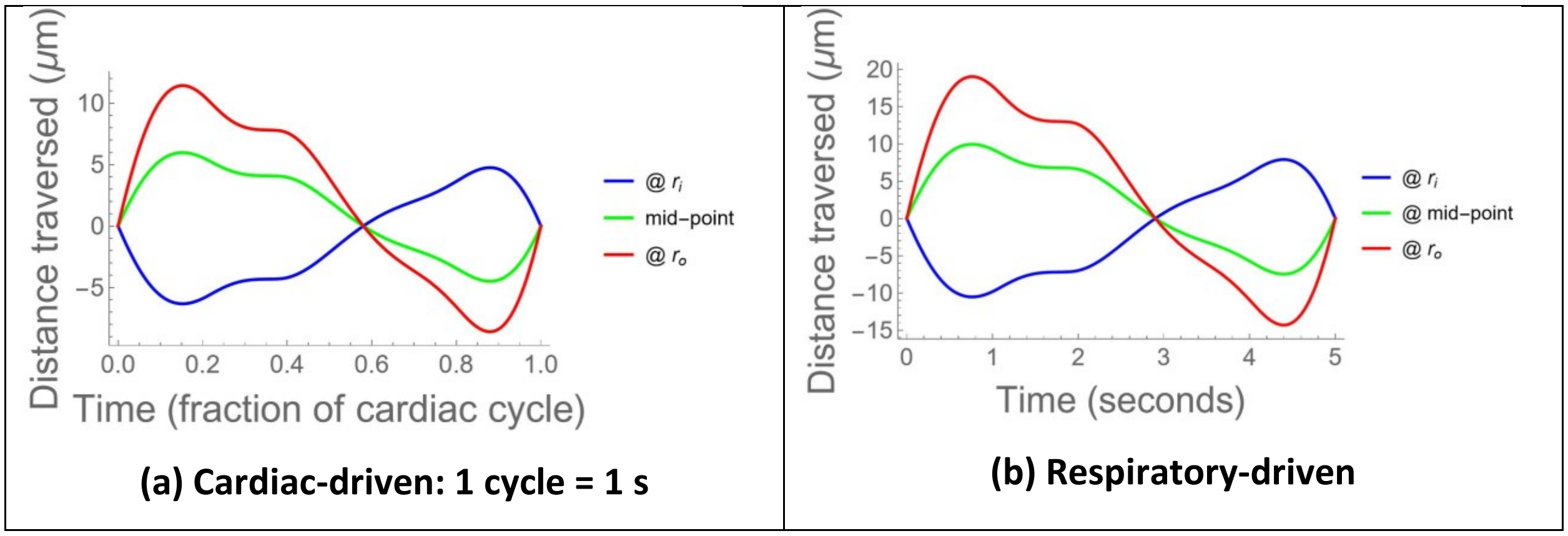

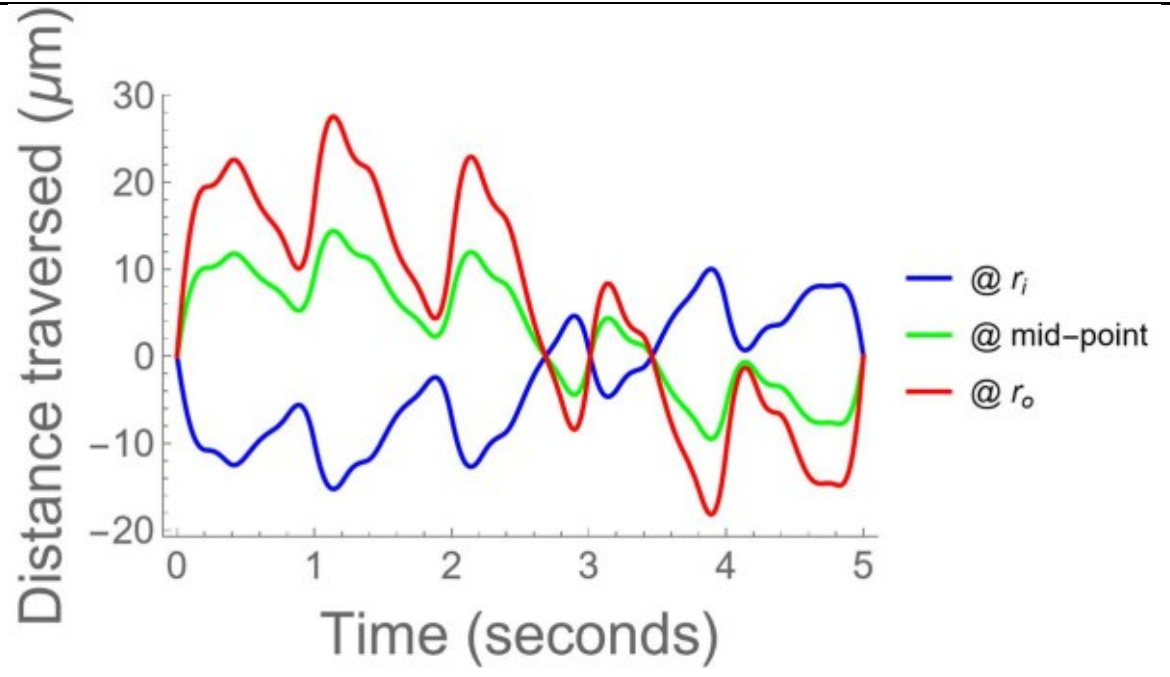

**(c) Addition of effects in-phase**

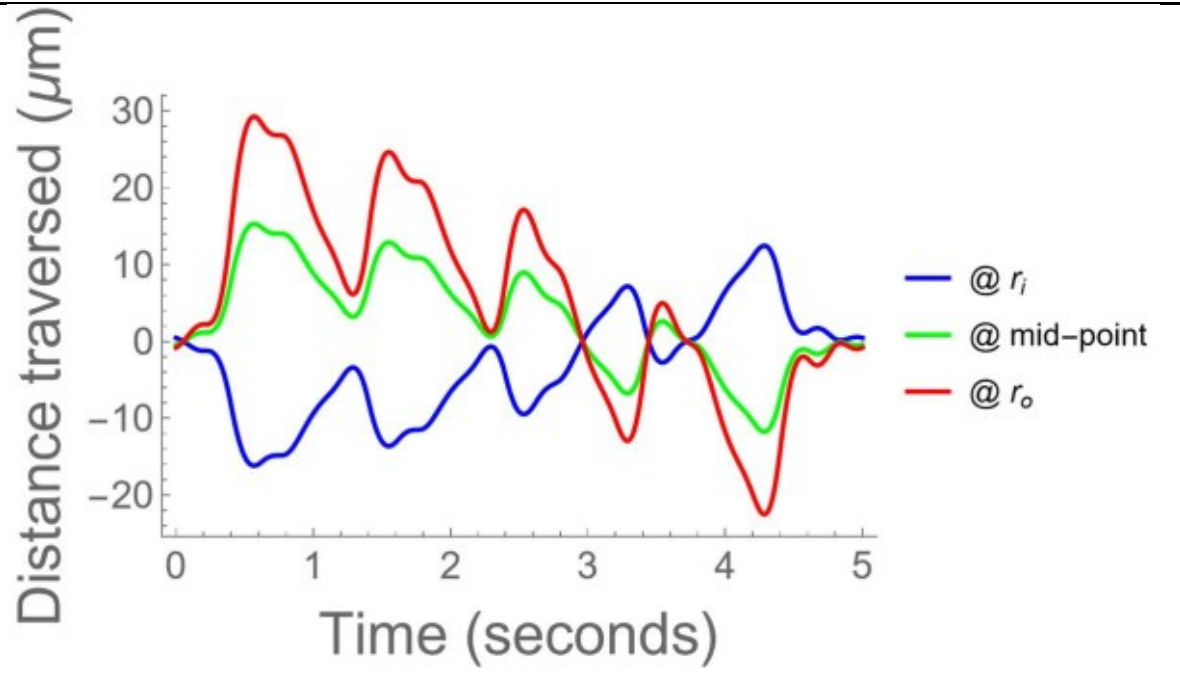

**4(d) Addition of effects out-of-phase**

In Figures 6(c) and (d), the respiratory effect has been added to that of the cardiac ones. Since the cardiac phase with respect to the respiratory phase is not known (and highly variable as well), Figure 6(c) shows what happens if both are started off at the same time. The cardiac cycle is 1 *s* and the respiratory cycle is 5 *s* in duration in our assumption. Both cycles are, thus, of constant as well as commensurate periods, so that the overall period is 5 s. This is, of course, an oversimplification, and the reality is that the flows will be more complex. Figure 6(d) shows this when we start the respiratory cycle at the point where the cardiac cycle is taking the particle past the starting point in the opposite direction, which is about 0.6 of the cardiac cycle. The calculation here has simplified this to be at the halfway point, ignoring the asymmetry of the cycles. It is clear that in both cases, the excursions are substantively greater than those for the cardiac cycle acting alone.

While these distances may seem rather high, the study of (Heukensfeldt Jansen, Abad et al. 2025) shows substantially greater interstitial velocities or distances were observed. They used a method based on phase-contrast imaging with the acronym SCIMI (see their paper for details) and provide color-coded speeds in, for example, Figure 6 of their paper. This is consistent with our order of magnitude calculations but unfortunately, their measurements involve total motion (tissue plus fluid). In contrast, the results of (Hirsch, Klatt et al. 2013) as described in their paper, yield the fluid velocity divergence as described above. The velocities then follow in the simple model we described. Of course, the inhomogeneous hydraulic properties of tissue will result in variations in these velocities. We do not now have much experimental authority for incorporating these. Indeed in unrelated work (Brady, Raghavan et al. 2020), we have shown that there is no significant difference between the hydraulic conductivity magnitudes in white and gray matter, though they are more directional in the former.

*3.2 Effective diffusivity due to advection in the oscillatory random walk model.*

To see if these results explain the higher dispersion found in the human tracer studies, we calculated the effective diffusion due to morphological dispersity. The nominal diffusivity of the tracer is 380 $\mu m^2/s$ whereas the observed diffusivity is 3.5 times that or about 1000 $\mu m^2/s$ (Sloots et al. 2021; Vinje et al. 2023). Figure 6 shows that an oscillatory walk with $N$ step and with each step being of length $d$ results in a mean square distance $Nd^2/3$. In the cardiac case, we see from Figure 4 that $d \approx 10$ μm four times over a period, so that in time $T$, the mean square distance is $T/\tau \times d^2/3$ where $\tau \approx 1/4$ s. Equating this with $6DT$ in three dimensions, where $D$ is the diffusivity, we get $D \approx 2d^2/9 \approx 20$ $\mu m^2/s$, which is insufficient. We note, however, that the recent experiments quoted at the end of Section 3.1 offer room for more optimism, but those results, nevertheless, will likely be insufficient. With the addition of respiration, as shown in Figure 4 bottom, we see several excursions of more than 20–40 μm and more for at least a portion of the cardiac half-cycle, which lead to effective diffusivities of between 80 and 300 $\mu m^2/s$. These rough calculations may easily be off by an order of magnitude, but the point is that morphological dispersion can contribute an effective diffusion at least as large as the nominal one. Moreover, for large molecules, which are of greater interest for therapeutic purposes, the nominal diffusivity is entirely negligible while the morphological dispersion remains unchanged *provided the advection remains the same* as mentioned in discussing Figure 1. The upper end of the calculated diffusivities (upon including the molecular diffusivity) is within a factor of 2 of the experimental data in (Vinje, Zapf et al. 2023). In any case, the calculated diffusivity is greater than the nominal diffusivity for even a small molecule. The effects will be more dramatic for larger molecules with sufficient advection, such as therapeutic protein molecules with molecular weights of 66 kDa (Brady, Raghavan et al. 2020) which continue to be explored for treating brain cancer and other brain diseases.  Such molecules have essentially the same advection as small-molecule gadolinium tracers with a molecular weight around 674 Da) and hence will display similar dispersivities due to this mechanism, though their molecular diffusivities are negligibly small to have a noticeable effect even in 24 hours.

## 4 Discussion and Conclusions

### *4.1 Background*

The impetus for the proposal in this paper was the work of Vinje and colleagues who measured the rates of flow of a CSF marker into and out of the human brain (Vinje, Ringstad et al. 2019) using the small MRI-visible molecule gadobutrol. They note that: "Our findings demonstrate that clinically-observed tracer influx and clearance patterns are compatible with (a) enhanced effective extracellular diffusion with $\alpha \approx 3.5$ combined with a local clearance rate of $r \approx 3.1 \times 10^{-3}$/min, or (b) extracellular diffusion augmented by advection with average fluid flow speeds of $|\varphi| \approx 1$–8 μm/min." In the quote the factor α is the factor multiplying the nominal or molecular diffusivity of the contrast reagent gadobutrol. In other words, they say that one possible explanation of their results is to discover a mechanism for augmenting the effective diffusivity. Note that the clearance patterns are observed over a period of 24 hours, and so the augmented diffusivity should be apparent over time scales of several hours, *ie,* of the order of $10^4 - 10^5 \times$ cardiac or respiratory cycle period.

This paper suggests a mechanism, first proposed here as far as we know, that would provide an additional effective diffusion coefficient required when tracer movement is observed over time much longer say $10^2 \times$ than a few seconds. *To avoid misunderstanding, we do not rule out alternative explanations of their experimental results.* For example, (Vinje et al. 2023) suggest in the alternative (b) in the quote above, that if the effective diffusivity remains the same then the movement could be accounted for by additional directional fluid convection. Such a convection would transport a molecule a distance that is, on average, proportional to time as opposed to the mechanism we suggest which has the hallmarks of a classical random walk where the displacement is proportional to the square root of the time elapsed.

While the influence of cardiac pulsations on the flow of cerebrospinal fluid (CSF) around the brain and the spinal cord and in the perivascular spaces has been observed and modeled for many years (for a review, see (Wagshul, Eide et al. 2011)), the effect that these pulsations could have on the interstitial flow fluid in the brain tissue has not been studied until recently (see (Lecchini-Visintini, Chung et al. 2024, Lecchini-Visintini, Chung et al. 2025)). The interstitial movement of molecules is of considerable importance for the delivery of proteins, viruses, and genetic factors, for the treatment of diseases of the central nervous system, and for pathological processes that might disrupt this flow. Considerable evidence (Syková and Nicholson 2008) has established that diffusion describes the movement of molecules over short distances, 100 to 200 μm, and relatively short times, seconds, and that the advection of fluid can be ignored because, if present, it is much slower. However, human MRI experiments using contrast reagent (chelated Gadolinium (Hingorani, Bernstein et al. 2015)) markers demonstrate that these molecules are spreading further than predicted by known diffusion or advection mechanisms, which indicates that some other factor must be involved. We suggest, based on the previous quantitative water flow measured by MRI in the human brain and our modeling, that a particular morphological dispersion (defined in this paper that more precisely below) produced by cardiac pulsations (as well as respiratory pulsations) is occurring and that this enhanced dispersion could help explain the more rapid and non-molecular weight-dependent movement of molecules in interstitial spaces. Furthermore, the rapid oscillation of fluid due to cardiac-induced pressure changes in the interstitial spaces may be important in the physical mixing of molecules and their interaction with cell surfaces. There have been a number of studies of CSF flow (Kelley and Thomas 2023) as well as advection and diffusion in PVS (Guo, Quirk et al. 2025) but the mechanism proposed here has been applied within the brain interstitium. Glymphatic flow has been discussed both theoretically (Tithof, Boster et al. 2022) and experimentally (Watts, Steinklein et al. 2019). High-field MRI has been used (Rey and Sarntinoranont 2018) followed up by a more extensive study on implications for perivascular transport (Rey, Farid et al. 2023). Note that Taylor–Aris dispersion is not considered in this paper because it is ineffective at the small scale of interstitial widths, as can be inferred from the calculations of Sharp and colleagues (Sharp, Carare et al. 2019), in particular extrapolations of Figure 5 of their paper.

### *4.2 Discussion*

The rapid cyclic movement in the interstitial spaces contrasts with the simple Brownian motion causing the diffusion of particles in a stagnant fluid space (Godin et al. 2017). The calculated dynamic oscillatory forces due to advective flow would be expected to move a molecule of neutral buoyancy over the length of the average distance from bifurcation to bifurcation in the densely packed gray matter within a half-period of either the cardiac or the respiratory cycle. For simplicity, it has been assumed that cells in the brain have a diameter of about 10 μm. The average size of a cell is approximately 1000 $\mu m^3$ with a range from 0.3 to 30 μm for a side if the cell were a cube. (Bundgaard, Regeur et al. 2001) Obviously, different types of gray matter have different cell sizes, and there is great heterogeneity in any given region. For computations, we found it necessary to make an approximation based on an average. This distance is within the range needed to produce morphological dispersion as we have defined it in this paper. Such an effect would clearly influence the measured effective diffusion.

Furthermore, since morphological dispersion is caused by a fluid convecting the molecules with it, the size of the molecule is not important as it is in pure diffusion. Many biologically important molecules are large (neurotrophins, beta amyloid, tau, and synuclein) and their movement by diffusion would be very limited, whereas morphological dispersion treats small and large molecules in the same way. The early brain tissue injection experiments found that the concentrations of large and small molecules decrease at the same rate as they are removed over time (Nicholson 2001). Thus, interstitial fluid movement could be the key to understanding the movement in gray matter of large biologically active molecules, medications, genetic constructs, and viruses.

Cardiac pulsations are likely to produce morphological dispersion, but respiration is likely to also play an important role. The influence of respiration is well documented in oscillatory CSF flow through the aqueduct of Sylvius, as measured in MRI studies that synchronize respiratory cycles with flow patterns. In human studies, respiration cycles move 4 to 5 times the volume per cycle in the aqueduct compared to cardiac pulsations (Vinje, Ringstad et al. 2019). This is because the respiratory effect is over a longer cycle time and not because of a bigger pressure gradient. When respiration prolongs flow in the large subarachnoid and ventricular spaces, it may well prolong fluid movements in interstitial spaces and produce much greater morphological dispersion. That this is likely is shown by the experiments by (Hirsch, Klatt et al. 2013). In those experiments, participants contracted their abdominal muscles during scan acquisition, and this maneuver increased the measured div($\boldsymbol{v}$) from $1.5 \pm 0.3$ to $2.2 \pm 0.4$ mL/mL/sec or about a 50% increase.

There is another issue we wish to consider. In Table 2 of (Vinje, Ringstad et al. 2019), the average of the peak oscillatory flow rate in the aqueduct due to cardiac pulsation is listed as 0.31 cc/s. Our calculations described above also yield a flow rate into the ventricles: Assuming this to be transferred via the aqueduct, the flow rate we get at the epoch of the peak in the oscillatory divergence from the velocities computed in Section 2.3 is 0.33 cc/s, which is surprisingly close.

In calculating respiratory effects, we have followed (Sloots, Biessels et al. 2021) in scaling these by 1/3 of the cardiac value, as discussed. Nevertheless, Table 2 of (Vinje, Ringstad et al. 2019) shows that the aqueductal flow from respiratory effects is just as great (0.35 cc/s). Our model, therefore, is inconsistent with that figure for the respiratory influence. If we were to take the respiratory-induced divergence to be as great as the cardiac-induced divergence, the desired augmentation of the diffusivity would be more comfortably reached.

Our simple models do not capture the full complexity of interstitial fluid movements due to variations in the cellular packing configuration of the fluid spaces, distances, and widths as well as changes in physiological states, like sleep or tissue damage resulting from disease states. It is unlikely that MRI techniques measuring divergence will be improved to the point that they will be able to find the small differences that are likely to be present in heterogeneous gray matter. It is, however, possible that the techniques could differentiate white matter regions from gray (Sloots, Biessels et al. 2020, Sloots, Biessels et al. 2021). The values for the overall movement of fluids in the tissues calculated by our models account for Eide's group MRI kinetics, but there must be a much larger variation if smaller diverse areas are to be visualized (Ringstad, Valnes et al. 2018). As mentioned in Section 2.1, there are known variations in $\phi$, the interstitial volume fraction, in gray matter regions, due to both physiological and pathological conditions. The effect of these is easy to point out in theory: if div($\mathbf{v}$) remains unaltered, the interstitial velocities are inversely proportional to $\phi$. For example, if the interstitial space increases during sleep, the velocity will decrease by the same factor. However, since we do not understand fully the physiology producing these oscillatory velocities, we are not in a position to assert that the tissue-averaged fluxes will remain the same.

One way to measure brain elasticity in vivo is to apply extracranial pulses produced by ultrasound to the brain. As a part of their paper, Hirsch did this experiment on the same participants using 25-Hz pulses (Hirsch, Klatt et al. 2013). The calculated elasticity was the same as that calculated with the cardiac gated method. The externally induced pressure waves worked in the same way as the cardiac pulse of blood into the brain. We hypothesize that the flow of molecules within brain tissue could be significantly enhanced by external driving forces with the correct magnitude (see coda below). Drug delivery from the CSF and movement through the tissue might be controllable. This delivery could be enhanced by timing the forces to the cardiac pulse or by inhibiting the flow with out-of-phase pulses. Our calculations based on Hirsch's experimental results make it clear that perturbations of the brain tissue from an energy source external to the skull could work. Such energy sources could be focused onto regional areas, as with ultrasound, to increase morphological dispersion. The safety and practicality of such interventions need to be investigated and, if successful, could be used to treat brain diseases by improving the distribution of drugs to targeted areas.

Whether or not such ways of implementing morphological dispersion work, our studies should change the way in which the movement of interstitial fluid is thought of. Cardiac pulsations as well as other factors, like respiration, rapidly and continually cause fluid to move in the gray matter of the brain and affect many important aspects of its physiology.

### *4.3 Coda: Inducing dispersivity*

To conclude with yet another speculative section, the range of frequencies and intensities of oscillations within the parenchyma that could induce dispersivity to enhance drug distribution in the brain by artificial means is evaluated. The standard bound the U.S. Food and Drug Administration requires for safety for vibrational excitation in the brain is that the so-called mechanical index (MI) (Nowicki 2020) be smaller than 1.9. For numerical convenience here, the safety number is taken to be 1. (The reference may be consulted as to how this number is arrived at, but avoiding the danger of cavitation bubbles of larger size is a major consideration. For brevity, we do not discuss safety limits for heating the tissue, quantified by a thermal index, *loc cit.*) We measure pressure in pascals and frequencies in hertz. Denoting the pressure and the period (=1/frequency) in these units as $P$ and $T$ (=1/$f$), the MI is

$$\mathrm{MI} := \frac{\text{pressure in MPa}}{\sqrt{\text{frequency in MHz}}} = 10^{-3}\frac{P}{\sqrt{f}} \tag{8}$$

Taking the mean velocity to be

$$\frac{\text{pressure}}{\text{density of fluid} \times \text{speed of sound}}, \tag{9}$$

and taking the density of the fluid to be that of water (1 gm/cc), the excursion

$$\begin{aligned} \text{distance in μm} &= 10^6 \times \text{distance in m} \\ &= 10^6 \times \frac{\text{pressure in Pa}}{1000\ \text{kg/m}^3 \times 1.55 \times 10^3} \times \frac{1}{2\ \times \text{frequency in Hz}} \\ &\approx 32 \times 10^{-2}\frac{P}{f} \end{aligned} \tag{10}$$

Again for numerical convenience the minimum distance is taken to be 32 μm, considerably larger than the 10 μm cellular widths. The expression in (8) must be less than unity for safety and that in (10) greater than 32 to allow more than enough excursion for molecules to spread according to the model described in the main sections, giving us the requirements:

$$10^{-3}\frac{P}{\sqrt{f}}<1;\quad 10^{-2}\frac{P}{f}>1 \tag{11}$$

whose solutions lie in the range

$$100f < P < 1000\sqrt{f} \tag{12}$$

Assuming that anything that is audible is automatically unacceptable and that periods greater than 10 s are also unacceptable, we plot the region below 20 but above 0.1 Hz as the yellow region in Fig 8.

**Figure 7**

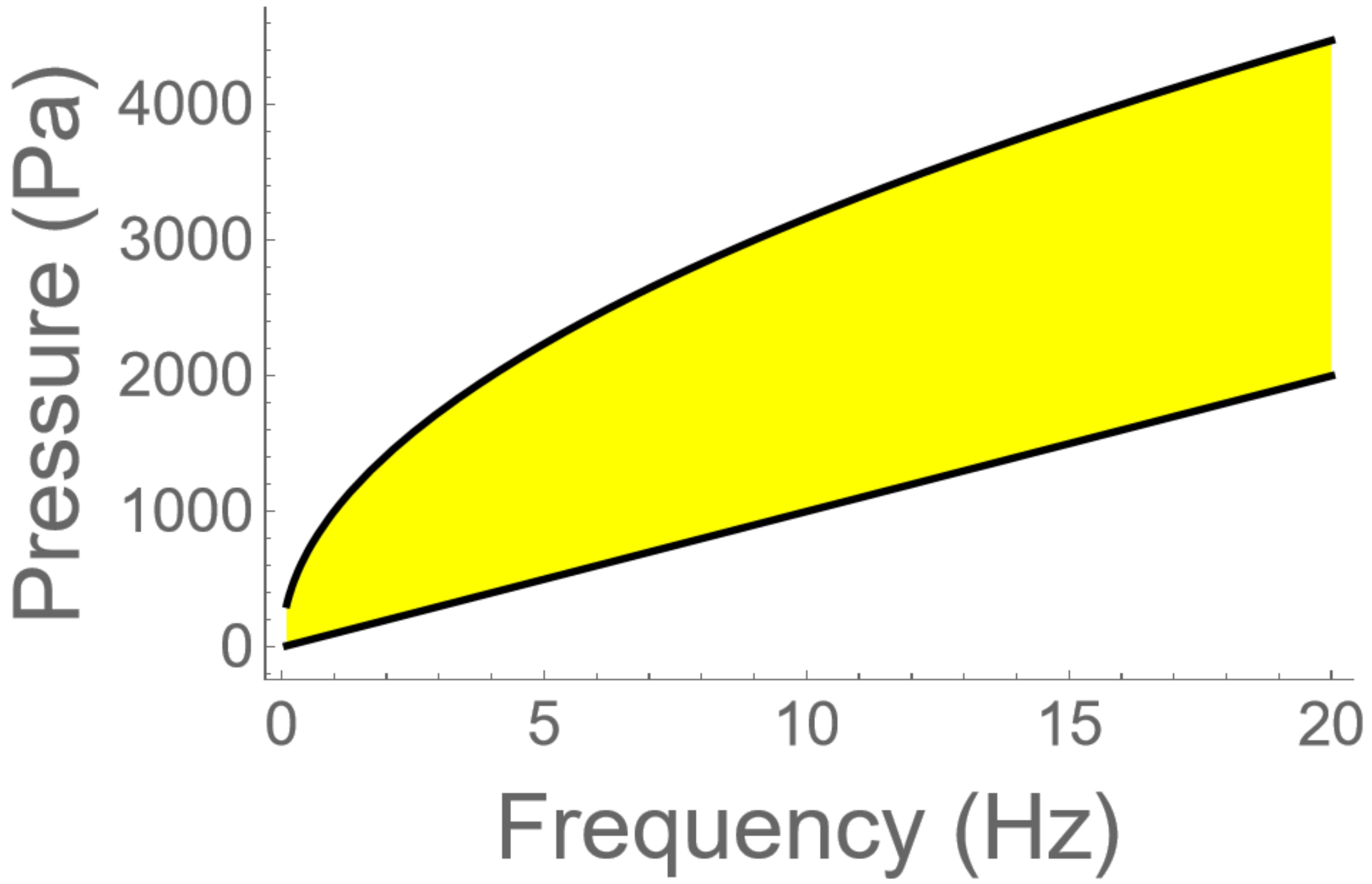

There are several serious problems to be considered here. Infrasound, while not audible, can cause severe adverse effects in humans (Møller 1984, Landstrom and Pelmear 1993). Moreover, the approximation that morphological dispersion – as we have used it – is equivalent to diffusive transport is valid only in the limit of long times. However, these times may be clinically unacceptable. Whether a suitable solution can be found, for example, by exploiting nonlinear phenomena such as in parametric arrays (Crocker 1998) or other methods to suppress standing waves (Tang and Clement 2009), remains a subject for the future. Other interventions could enhance advection and diffusion by different means not dependent on the mechanical dispersion stemming from molecules following different pathways in tissue. Such methods include ultrasound-induced streaming, for example, (Raghavan 2018, Boroumand, Mehrarya et al. 2022, Price, Hansen et al. 2023), and acoustically augmented transport (Yoo, Kim et al. 2023). Two high-frequency waves with a very small difference frequency, namely a high-frequency carrier with slow amplitude modulation, can cause *oscillatory* streaming, which would be a novel application (see (Marston 1980) for reference to such a phenomenon). This would be in addition to the steady streaming caused by the attenuation of the high-frequency carrier wave.

## Figure captions

**Fig 1. Illustration of dispersion due to the multiple pathways in an oscillatory flow.**

Our approach is based on considering two separated spatial and time scales. We use "milliscopic" (in the millimeter range of spatial scales) MRI measurements of movements over a time scale of 1 s to infer the bulk flow of fluid in the interstitium. We then modify the flow and the transport of particles advected by the flows based on microscopic (micron-scale) considerations. Later, we will examine the consequences for large spatial scales (centimeters) and long times (hours). In the first half-cycle of an oscillatory flow, it is *assumed* in the illustration that a particle moves *past a branch*. In the subsequent half-cycle, beginning when the particles reach the branch on their return trip, it is *assumed* that they will choose a branch randomly as long as it does not switch back. In the simulations we constrain the direction taken to have a positive projection on the direction approaching the branch but otherwise random. It can be seen that the particles have irreversibly drifted away from their starting point despite the macroscopic flows being fully reversible. There is no contradiction between these two statements.

**Fig 2. Random walk simulations.**

The figure compares 100 steps of two random walk samples (with an additional first step that was chosen to be the same for both). The steps in red are for an oscillatory walk, while those in green are for a purely random walk. This figure illustrates clearly (albeit for one sample each) that the steady walk wanders further than the oscillatory one. Numerical simulations show that the root mean square distance obtained for the oscillatory walk is a factor $1/\sqrt{3}$ of the corresponding root mean square distance for the steady velocity walk (Figure 3).

**Fig 3. Number of steps in a random walk versus mean square distance.**

The simulations in all cases have variations smaller than the size of the dots used to show the results. For the largest number of steps (1000), the results shown are over a sample size of 4000 runs. Shorter walks have larger sample sizes. Each step was unit length. For the fully random walk (blue dots), the mean square distance is equal to the number of steps, which is a well-known exact result for a fully random walk. For the oscillatory walk (red dots), the mean square distance is 1/3 times the number of steps according to the numerical simulations.

Obviously, if each step length is fixed and is of length $d$, the corresponding mean square distance will be $d^2$ as large.

**Fig 4. Water movement over the cardiac cycle.**

Water movement in brain tissue is represented by the divergence div($\mathbf{v}$). The values are from (Hirsch, Klatt et al. 2013) and were obtained by MRI-gated cardiac imaging.

**Fig 5. Diagram of the model used for computations.**

The CSF is represented in blue in the cranial subarachnoid space (SAS) and ventricle. The subscripts denote the transition from one region to the next. $r_i$ is the radius at the ventricular/inner gray matter boundary, $r_{gw}$ is the radius at the inner gray matter/white matter boundary, $r_{wc}$ is the radius at the white matter/cortical gray matter boundary, and $r_o$ is the cortical/SAS interface radius. In the model used for the flows, the different tissue regions have exactly the same hydraulic properties, so that the geometry of the region boundaries is used only when we plot the distances of particles carried by the macroscopic flow at different places.

**Fig 6. Displacement of a particle advected by macroscopic velocities.**

Shown are the results of computing the displacements of a particle advected by the macroscopic velocities, as given in Section 2.3.2, when the particle commences its displacement at the beginning of a cycle at the positions noted by the legends in the figure. Namely, we show the displacements for a particle near the inner boundary (ventricles), in the middle of the parenchyma, and near the outer boundary (SAS). Figure 6a shows this for the cardiac cycle, while Figure 6b shows it for the respiratory cycle, with the scaling assumptions discussed in Section 2.3.2. In Figures 6c and 6d, we have added the two effects: in 6c, the cardiac and respiratory cycles start together while in Figure 6d, they are in opposition.

**Fig 7. Pressure and frequency ranges allowed for the advective enhancement of movement.**

The yellow region shows the available region in a harmonic pressure excitation versus frequency plot that would be sufficient to move a particle advectively at least 32 μm (which is safely larger than the distance between branches; see text) and yet have an MI of less than 1 for physiological safety. See the text for an explanation and discussion.

## Declarations

*Ethics approval and consent to participate* Not applicable

*Consent for publication* Not applicable.

*Availability of data and materials* The data for the calculations were taken from the referenced papers. The Mathematica calculations can be obtained from the corresponding author on reasonable request.

*Competing interests* The authors declare that they have no competing interests.

*Funding* None

*Authors' contributions* The authors worked cooperatively on the paper and its basic ideas. RR was responsible for the mathematical calculations and the idea of applying morphological diffusion to the process. RP worked on the physiological aspects.

*Acknowledgements* We are very appreciative of the time Dieter Klatt at UIC took in answering questions about his measurements of water flow in the brain using MRI methods. We also would like to thank Jeffrey Iliff, Jeffrey Heys, and Lori Ray for reviewing an earlier version of our manuscript and providing excellent comments.

*Authors' information*

Raghu Raghavan. President, Therataxis, LLC, Baltimore, MD.
Richard D Penn MD Adjunct Professor Biomedical Engineering UIC Chicago, Illinois and Emeritus Professor of Neurosurgery Rush Medical School Chicago, Illinois

## References (EndNote)